\documentclass[aps,prb,twocolumn,showpacs,preprintnumbers,amsmath,amssymb,superscriptaddress]{revtex4-1}
 
\usepackage{graphicx}
\usepackage{dcolumn}
\usepackage{bm}
\usepackage{longtable}
\usepackage{color}
\usepackage{textcomp}
\usepackage[utf8]{inputenc} 
\usepackage{hyperref}  
\usepackage{gensymb}   

\begin{document}

\newcommand{\changeStart}{\color{black}}
\newcommand{\changeStop}{\color{black}}

\preprint{PREPRINT (\today)}

\title{Magnetic torque study of the temperature dependent anisotropy parameter in overdoped superconducting single-crystal YBa$_2$Cu$_3$O$_7$ }
 
\author{S.~Bosma}
\email{sbosma@physik.uzh.ch}
\affiliation{Physik-Institut der Universit\"at Z\"urich, Winterthurerstrasse 190, CH-8057 Z\"urich, Switzerland}

\author{S.~Weyeneth}
\affiliation{Physik-Institut der Universit\"at Z\"urich, Winterthurerstrasse 190, CH-8057 Z\"urich, Switzerland}

\author{R.~Puzniak}
\affiliation{Institute of Physics, Polish Academy of Sciences, Aleja Lotnik\'ow 32/46, PL-02-668 Warsaw, Poland}

\author{A.~Erb}
\affiliation{Walther Meissner Institut, Bayerische Akademie der Wissenschaften, D-85748 Garching, Germany}

\author{A.~Schilling}
\affiliation{Physik-Institut der Universit\"at Z\"urich, Winterthurerstrasse 190, CH-8057 Z\"urich, Switzerland}

\author{H.~Keller}
\affiliation{Physik-Institut der Universit\"at Z\"urich, Winterthurerstrasse 190, CH-8057 Z\"urich, Switzerland}

\begin{abstract}

An overdoped YBa$_2$Cu$_3$O$_7$ single crystal was studied by SQUID and torque magnetometry in order to investigate the temperature dependence of the anisotropy parameter close to the transition temperature $T_{\rm c}$ ($0.87~T_{\rm c} < T < T_{\rm c}$). Angle dependent torque measurements were performed and analyzed with the widely used Kogan model [Phys. Rev. B {\bf 38}, 7049 (1988)] as well as with an extended model by Hao and Clem [Phys. Rev. Lett. {\bf 67}, 2371 (1991)], taking into account the influence of the vortex cores on the magnetization. Both approaches yield similar results, with an out-of-plane anisotropy parameter around 6.5 which slightly increases with decreasing temperature, and a temperature independent in-plane anisotropy parameter $\gamma_{ab} =  1.12(5)$.

\end{abstract}
  
\pacs{74.20.De; 74.25.Ha; 74.72.-h}

\maketitle

\section{Introduction}

Since the discovery of high-temperature superconductivity in the cuprates,\cite{Bednorz1986} the anisotropic magnetic properties of layered superconductors were extensively studied (see {\it e.g.} \changeStart Refs. \onlinecite{Farrell1988, Zech1996, Willemin1999, Hofer2000, Kohout2007a, Angst2002, Kortyka2010, Kortyka2010a}\changeStop). All superconductors with a transition temperature $T_{\rm c} > 30~\rm{K}$ have a layered structure. In particular, the superconducting gap of cuprates was found to be strongly anisotropic due to the crystal structure consisting of weakly coupled superconducting CuO$_2$ planes.\cite{Campuzano1990} It is interesting to investigate how the anisotropic properties change as a function of \changeStart thermodynamic parameters and \changeStop doping within a particular family of cuprates, and to explore their common features by comparing various families. 

The gap structure can be probed directly by exciting superconducting carriers. Importantly, the energy needed for this, {\it i.e.} the energy gap, may be accessed by probing the magnetic penetration depth $\lambda$.\cite{Chandrasekhar1993} In a layered superconductor the gap structure is strongly anisotropic, thus $\lambda$ is anisotropic as well. \changeStart The magnetic penetration depth related to a supercurrent flowing along the $i$-axis ($i = a, b, c$) is denoted as $\lambda_i$,  and the penetration depth anisotropy between two crystallographic directions $i$ and $j$ is $\gamma_{ij} = \lambda_i$/$\lambda_j$. \changeStop 

In the anisotropic Ginzburg-Landau theory, which is the most commonly applied phenomenological description of layered superconductors, the anisotropy is described by the temperature independent effective mass anisotropy (assuming $\lambda_i$/$\lambda_j = (m^*_i$/$m^*_j)^{1/2} = H^{||j}_{c2}$/$H^{||i}_{c2}$, where $H^{||k}_{c2}$ is the upper critical field along the $k$-axis). However, a temperature dependent anisotropy was observed in various superconductors, especially in the two-gap superconductor MgB$_2$ (see Refs.~\onlinecite{Nagamatsu2001, Angst2002}), and was explained as a consequence of the presence of two superconducting gaps. A similar temperature dependence was also observed in iron-based superconductors,\cite{Weyeneth2009a} in which evidence for two-band superconductivity was provided by several experiments, including point contact spectroscopy\cite{Shan2008, Wang2009} and ARPES.\cite{Ding2008, Zhao2008} Multi-gap superconductivity seems to be more common than first expected, as indications of it were also observed in cuprates.\cite{Khasanov2007, Khasanov2007a, Khasanov2008} It may be related to the temperature dependence of the anisotropy,\cite{Dahm2003} as in the case of MgB$_2$\cite{Angst2002} and the iron-based superconductors.\cite{Weyeneth2009a} However, there may be other reasons for this temperature dependence: the anisotropy of the gap,\cite{Chandrasekhar1993, Kogan2002} the anisotropy of the Fermi surface,\cite{Butler1980} or strong coupling.\cite{Werthamer1967} A temperature dependent anisotropy parameter was also observed in cuprates (see {\it e.g.} Refs.~\onlinecite{Shibauchi1994, Hosseini1998, Khasanov2007, Kortyka2010a}). This rises the question whether the temperature dependence of the anisotropy is a common property of all layered high-$T_c$ superconductors, and how it is linked to the gap structure.

A recent study of the cuprate superconductor SmBa$_2$Cu$_3$O$_{7-\delta}$ facing this question was limited to the underdoped region only.\cite{Kortyka2010a} It was shown that the temperature dependence of the anisotropy is more pronounced for samples with lower oxygen content. Such samples are characterized by a well developed pseudogap, {\it i.e.} an additional energy scale which may play a similar role in the development of the temperature dependence of the anisotropy as the multi-gap structure in MgB$_2$ and iron-based superconductors. Therefore, it is very important to perform reliable studies of the temperature dependence of the penetration depth anisotropy for optimally doped and overdoped cuprates. In this doping range the pseudogap vanishes or eventually overlaps with the superconducting gap.

Taking all of the above into account, we decided to study the temperature dependence of the anisotropy of a detwinned, almost fully oxygenated, overdoped single crystal of YBa$_2$Cu$_3$O$_{7-\delta}$. This system exhibits an anisotropic energy gap,\cite{Kirtley2006} and several experiments indicate an order parameter of {\it s}+{\it d} wave symmetry.\cite{Smilde2005, Khasanov2007}
 
Here, we report on torque measurements of the anisotropy parameter of an overdoped YBa$_2$Cu$_3$O$_7$ single crystal. Torque magnetometry provides a direct method to study the anisotropic magnetic properties of superconductors, contrary to methods measuring physical quantities separately along different crystallographic directions from which the anisotropy parameter is determined. An analytical approach for the analysis of experimental data based on the solution of the Hao-Clem\cite{Hao1991a} functional is applied, which allows to investigate anisotropic extreme type-II superconductors beyond the London approximation. For simplicity, the London approximation of the anisotropic Ginzburg-Landau theory (AGLT),\cite{Ginzburg1950, Tinkham1996} in which simplifications of the geometry of the vortex structure are made, is often used for analyzing experimental data. However, as discussed by Hao and Clem\cite{Hao1991a} this approximation may not necessarily be adapted to the interpretation of magnetization measurements, and thus both approaches are compared in this work.

Section II gives a brief review of the London and of the Hao-Clem models in connection with the torque magnetometry technique used in this work. The experimental details are described in Sec.~III. The results and the discussion are presented in Sec.~IV, followed by the conclusions in Sec.~V.

\section{London and Hao-Clem models}

The angular dependent magnetization $\overrightarrow{M}$ of a sample with volume $V$ and magnetic moment $\overrightarrow{m}$ is derived from the free energy $F$ of an anisotropic superconductor in the mixed state\cite{Kogan1988}
\begin{equation}
\overrightarrow{M}(\theta,H)=\frac{\overrightarrow{m}(\theta,H)}{V}=-\frac{1}{V}\vec{\nabla}_{B}F.
\end{equation}
The magnetic torque
\begin{equation}
\vec{\tau}=-\vec{\nabla}_{\theta}F=\mu_0V\left(\vec{M}\times\vec{H}\right)
\label{torqueeq}
\end{equation}
is related to $\overrightarrow{M}$ and the angle $\theta$ between $\overrightarrow{H}$ and the crystallographic $c$-axis.
 
A direct calculation of $F$ within AGLT is not trivial, since $F$ depends on the exact distribution of vortices and thus on the local magnetic induction $B(H)$ inside the superconductor. However, $F$ can be expressed within the so-called London limit, assuming that the influence of the finite vortex core size can be neglected. This is valid if the vortex core size is very small compared to the vortex itself, \textit{i.e.} the penetration depth is much larger than the coherence length. Anisotropic superconductors exhibit disctinct magnetic properties along the principal axes $a, b$, and $c$. In layered superconductors the largest anisotropy is observed between the $c$-axis and the layers ($ab$-plane). Therefore we may approximate the orthorhombic structure of YBa$_2$Cu$_3$O$_7$ by a tetragonal one,\footnote{As indicated in Ref.~\onlinecite{Kogan1988}, YBa$_2$Cu$_3$O$_7$ is the typical structure where this uniaxial approximation can be made} introducing the anisotropy parameter 
\begin{equation}
\gamma=\frac{\lambda_c}{\lambda_{ab}},
\end{equation}
where the in-plane magnetic penetration depth ${\lambda_{ab}} = \sqrt{\lambda_a\lambda_b}$. The magnetization $M$ and the torque $\tau$ are derived in the so-called Kogan model\cite{Kogan1988} as
\begin{equation}\label{ML}
M_{\rm L}(\theta,H)=-\frac{\Phi_0\epsilon(\theta)}{8\pi\mu_0\lambda_{ab}^2}\ln\left(\frac{\eta H_\mathrm{c2}^{||c}}{\epsilon(\theta)H}\right)
\end{equation}
and
\begin{equation}\label{torqueL}
\tau_{\rm L}(\theta,H)=-\frac{V\Phi_0H}{16\pi\lambda_{ab}^2}\left(1-\frac{1}{\gamma^2}\right)\cdot\frac{\sin(2\theta)}{\epsilon(\theta)}\ln\left(\frac{\eta H_{\rm c2}^{||c}}{\epsilon(\theta)H}\right).
\end{equation}
Here, the index L indicates the London approach, $\Phi_0$ is the magnetic flux quantum, and $\epsilon(\theta)$ is the angular scaling function
\begin{equation}
\epsilon(\theta)=\sqrt{\cos^2(\theta)+\frac{1}{\gamma^2}\sin^2(\theta)}.
\end{equation}
The parameter $\eta$ accounts for uncertainties due to the approximation of the London limit ({\it e.g.} the neglected suppression of the order parameter inside the vortex cores).

Hao and Clem\cite{Hao1991a} showed by \changeStart analyzing \changeStop the free energy within AGLT that the parameter $\eta$ cannot be constant in the entire magnetic field range $H_{\rm c1}<H<H_{\rm c2}$, \changeStart which is also evident from more recent theoretical work\cite{Pogosov2001} beyond the Hao-Clem model. \changeStop The correct functional form of $F$ by Hao and Clem\cite{Hao1991a, Hao1991b, Hao1991} incorporates in the expression for $M$ and $\tau$ the empirical functions $\alpha(h)$ and $\beta(h)$, where $h$ denotes the reduced field
\begin{equation}
\label{htheta}
h(\theta)=\frac{H}{H_\mathrm{c2}(\theta)}.
\end{equation}

Their generalized treatment of the mixed state of a superconductor, which includes the vortex core contribution to the free energy functional, yields a more realistic formula for the magnetization\cite{Hao1991a}
\begin{equation}
\label{MHC}
M_{\rm HC}(\theta,H)=-\alpha(h(\theta))\frac{\Phi_0\epsilon(\theta)}{8\pi\mu_0\lambda_{ab}^2}\ln\left(\frac{\beta(h(\theta))}{h(\theta)}\right),
\end{equation}
where the index HC indicates the Hao-Clem model. According to Eq.~(\ref{torqueeq}), the torque is written as
\begin{eqnarray}
\label{torqueHC}
\tau_{\rm HC}(\theta,H)	&=&		-\alpha(h(\theta))\frac{V\Phi_0H}{16\pi\lambda_{ab}^2}\left(1-\frac{1}{\gamma^2}\right)\\\nonumber
&\cdot&\frac{\sin(2\theta)}{\epsilon(\theta)}\ln\left(\frac{\beta(h(\theta))}{h(\theta)}\right).
\end{eqnarray}

Here, taking into account the suppression of the order parameter in the vortex core leads to a modification of Eqs.~(\ref{ML}) and (\ref{torqueL}) by including the functions $\alpha(h)$ and $\beta(h)$. These functions account for the correction of the in-plane magnetic penetration depth $\lambda_{ab}$ and the $c$-axis upper critical field $H_\mathrm{c2}^{||c}$, respectively. For $\alpha(h) = 1$ and $\beta(h) = \eta$, Eqs.~(\ref{MHC}) and (\ref{torqueHC}) reduce to  Eqs.~(\ref{ML}) and (\ref{torqueL}) of the London limit. Within the HC treatment no analytical formulas for $\alpha(h)$ and $\beta(h)$ can be derived easily. However, for a Ginzburg-Landau parameter $\kappa>>1$ the following values for $\alpha$ and $\beta$ are found\cite{Hao1991a}

\begin{eqnarray}
\label{HCestim}
0.02\lesssim h\lesssim0.1 &:& \alpha(h)\simeq0.84,~ \beta(h)\simeq1.08 \\
0.1\lesssim h\lesssim0.3 &:& \alpha(h)\simeq0.70,~ \beta(h)\simeq1.74~.
\end{eqnarray}
 
 It is clear that although $\alpha$ and $\beta$ are assumed to be constant in the London limit, they are field dependent and may vary considerably with magnetic field ($\alpha$ and $\beta$ are fully determined by the reduced field $h$).\cite{Hao1991a}

\begin{figure}[t!]
\centering
\vspace{-0cm}
\includegraphics[width=0.95\linewidth]{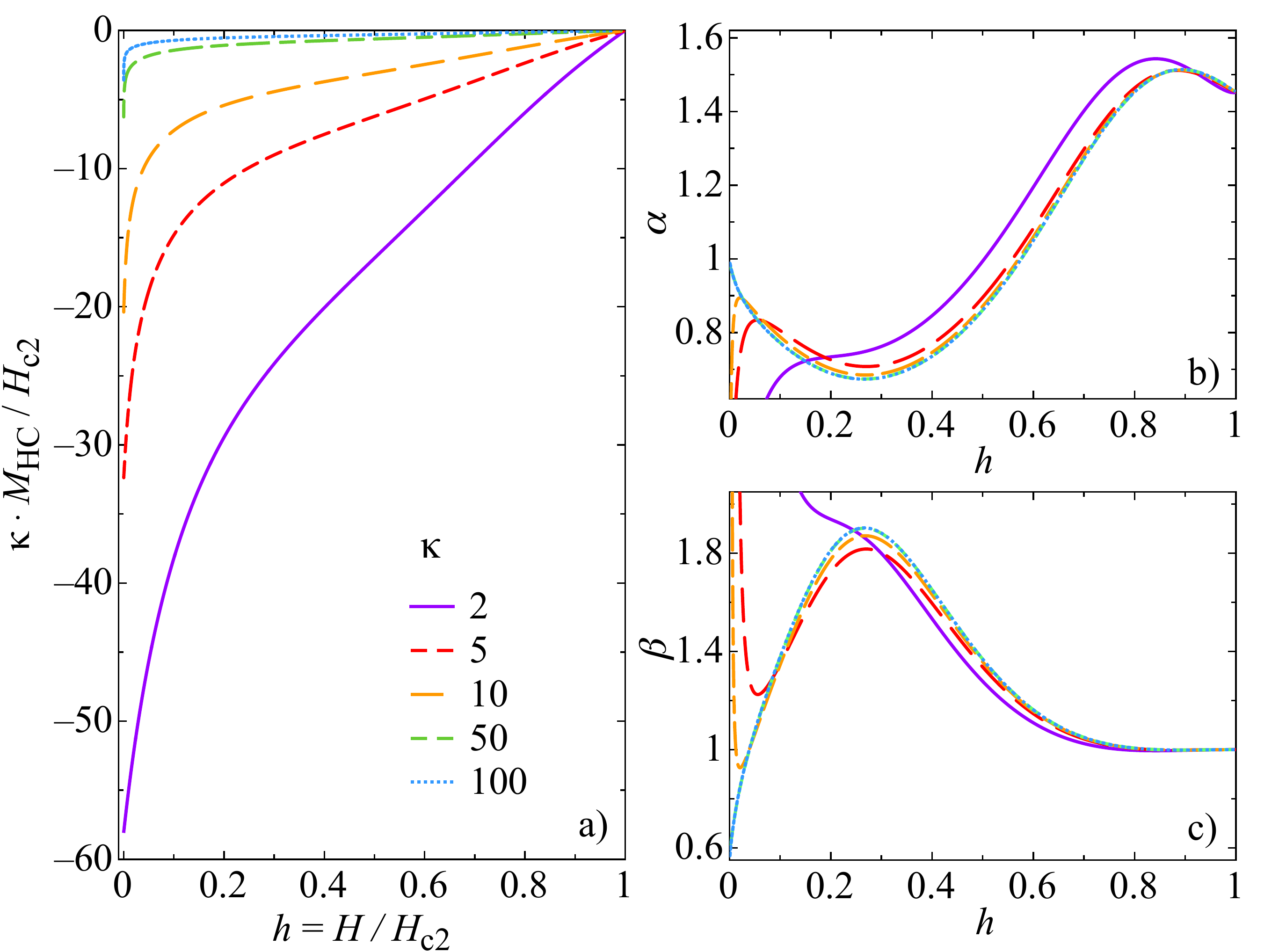}
\caption{(Color online) a) $\kappa M_{\rm HC}(h)/H_{\rm c2}$ for various values of $\kappa$ calculated using the HC model. b) Empirical parameter $\alpha(h)$ extracted from $M_{\rm HC}(h)/H_{\rm c2}$. c) Empirical parameter $\beta(h)$ extracted in the same way as $\alpha(h)$ in panel b). Both $\alpha(h)$ and $\beta(h)$ are essentially independent of $\kappa$ for $\kappa>50$.}
\label{MHC_calc}
\end{figure}

In Fig.~\ref{MHC_calc} we present the numerically calculated field dependence of the reduced magnetization $M_{\rm HC}(h)/H_{\rm c2}$ and compare it with the empirical Eq.~(\ref{MHC}) in order to extract $\alpha(h)$ and $\beta(h)$. The quantity $\kappa M_{\rm HC}(h)/H_{\rm c2}$ for $2<\kappa<200$ is presented in Fig.~\ref{MHC_calc}a (for clarity, $\kappa M_{\rm HC}(h)/H_{\rm c2}$ is shown instead of $M_{\rm HC}(h)/H_{\rm c2}$). Obviously, $M_{\rm HC}(h)/H_{\rm c2}$ strongly depends on $\kappa$. The functions $\alpha(h)$ and $\beta(h)$ are presented in panels b) and c), respectively. For $\kappa>50$, the functions $\alpha(h)$ and $\beta(h)$ become essentially independent of $\kappa$.  The derived $\alpha(h)$ and $\beta(h)$ are in good agreement with the values estimated by Hao and Clem\cite{Hao1991a} given in Eq.~(\ref{HCestim}). 

Analyzing magnetic torque experiments by means of the above described theoretical model by Hao and Clem, one should note that the parameter $\kappa$ is the isotropic Ginzburg-Landau parameter $\kappa = \lambda$/$\xi$, where $\xi$ is the coherence length, and $\lambda$ is the magnetic penetration depth. However, for a layered superconductor, the Ginzburg-Landau parameter is anisotropic: $\kappa$ has to be replaced by $\kappa(\theta) = \kappa_c$/$\epsilon(\theta)^2$, where $\kappa_{c} = \lambda_{c}$/$\xi_{c} = \gamma^2 \lambda_{ab}$/$\xi_{ab} = \gamma^2 \kappa_{ab}$. The functions $\alpha(h)$ and $\beta(h)$ then depend on the angle $\theta$ not only via $h(\theta)$, but also via $\kappa(\theta)$. For YBa$_2$Cu$_3$O$_7$ in a field of 1.4~T at $T = 80~\rm{K}$, $h(\theta)$ varies approximately between 0.1 ($\theta$~=~0\textdegree) and 0.01 ($\theta$~=~180\textdegree) when the field is turned from the $c$-axis to the $ab$-plane. In this case, $\alpha$ and $\beta$ strongly depend on $h(\theta)$ (see Figs.~\ref{MHC_calc}b and c), but for $\kappa(\theta)>50$ they are independent of $\kappa(\theta)$. Using the values for $\lambda_{ab}(T = 0)$ and $\xi_{ab}(T = 0)$ for YBa$_2$Cu$_3$O$_7$ from Refs.~\onlinecite{Kiefl2010, Welp1989}, one gets $\kappa_{ab} \simeq 400 $, which means that $\kappa(\theta)>400$ for all $\theta$. Thus, for YBa$_2$Cu$_3$O$_7$ the parameter $\kappa(\theta)$ has \changeStart negligible \changeStop influence on $\alpha$ and $\beta$. As a consequence, \changeStart taking into account \changeStop the anisotropy on the Ginzburg-Landau parameter $\kappa$ in the HC model does not lead to a more reliable determination of $\alpha$ and $\beta$.

\section{Experimental details}

\begin{figure}[t!] 
\centering
\vspace{-0cm}
\includegraphics[width=0.7\linewidth]{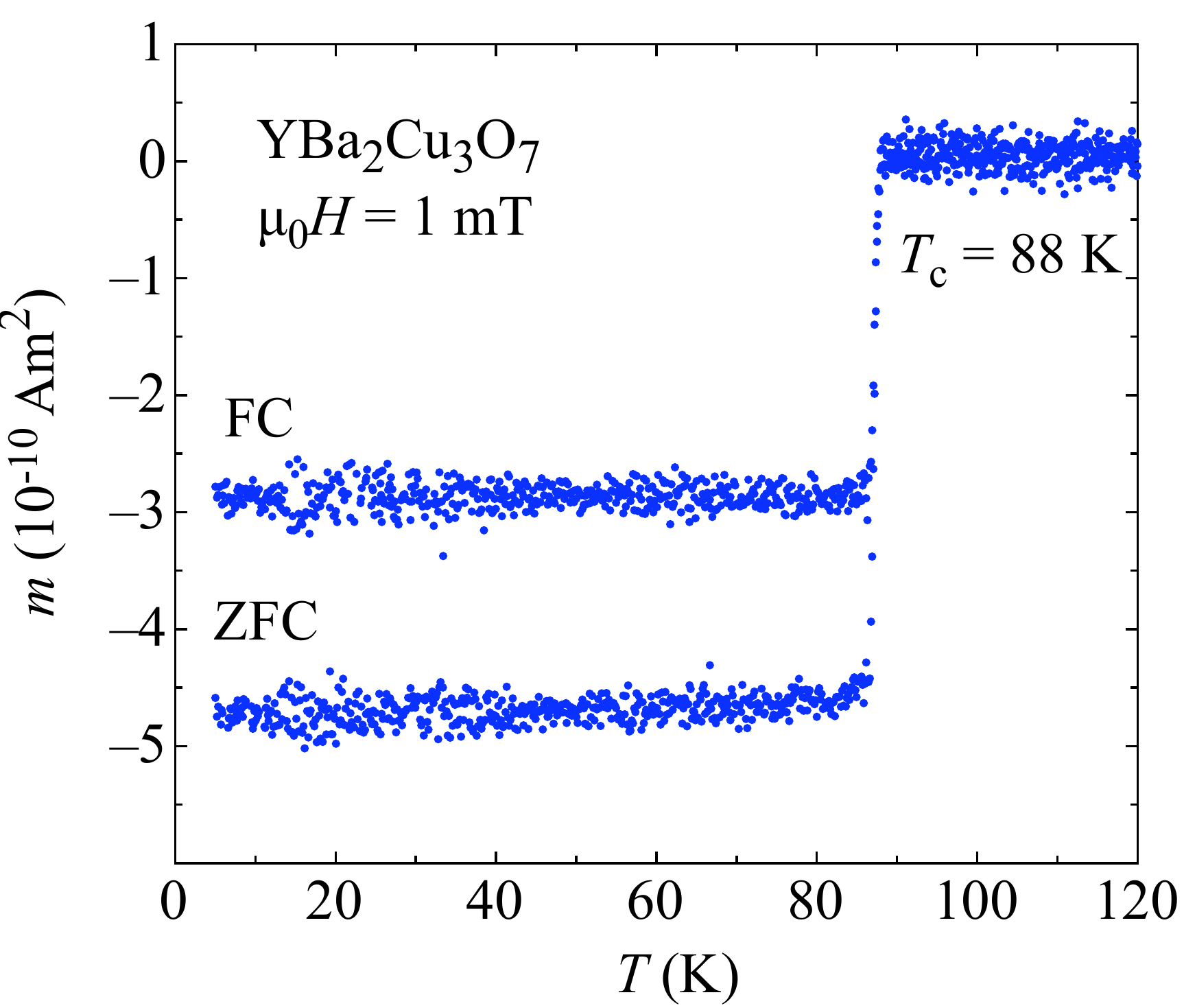}
\caption{(Color online) Magnetic moment $m(T)$ of the YBa$_2$Cu$_3$O$_7$ single crystal, measured in a magnetic field $\mu_0H=1$~mT parallel to the $ab$-plane. In the \changeStart ZFC \changeStop mode the field is applied once the sample is cold, whereas in the \changeStart FC \changeStop mode the sample is cooled while the field is applied. The \changeStart ZFC \changeStop and \changeStart FC \changeStop magnetization curves show a sharp transition with a transition temperature $T_{\rm c}$~=~88 K, indicating a high quality of the crystal.}
\label{m_T}
\end{figure}

The sample studied is an overdoped detwinned single crystal YBa$_2$Cu$_3$O$_7$ grown in BaZrO$_3$ crucibles, with dimensions 130$\times$160$\times$50 $\mu$m$^3$ and $T_{\rm c} \simeq 88$ K. Crystal growth in BaZrO$_3$ yields samples of highest purity.\cite{Erb1996} To fully oxygenate the crystal a high pressure annealing was performed at 300~$\celsius$ in 100 bar of oxygen. Samples produced in this way show no anomalies, \changeStart{\it e.g.} fishtail effect,\cite{Erb1996a}\changeStop and have very low pinning.\cite{Erb1999} A {\it Quantum Design} MPMS XL SQUID magnetometer was used to determine $T_{\rm c}$ (Fig.~\ref{m_T}). The temperature dependence of the magnetic moment $m$ was measured in a small field $\mu_0H = 1$ mT parallel to the $ab$-plane in zero field cooled (\changeStart ZFC\changeStop) and field cooled (\changeStart FC\changeStop) mode. The small difference between $m(T)$ obtained in the two modes and the sharp transition observed indicate a good quality of the crystal.

The torque measurements were carried out using a home-made torque magnetometer.\cite{Kohout2007a} The piezoresistive sensor used consists of a platform connected to piezoresistive legs which are bent when the sample mounted on the platform undergoes a torque. The resulting resistance change in the piezoresistors is detected by a Wheatstone bridge. The read-out voltage is proportional to the torque magnitude $\tau$. The small dimensions needed for the sample allow the study of high-quality single crystals. 
 
\section{Results and discussion}

The torque measurements were performed in an external field of 1.4 T and 1 T in order to check for a possible field dependence of the anisotropy. Since the sample has diamagnetic and anisotropic properties, its magnetization $\overrightarrow{M}$ is not quite aligned with the field $\overrightarrow{H}$, which results in a torque $\overrightarrow{\tau} \propto \overrightarrow{M} \times \mu_0\overrightarrow{H}$ according to Eq.~(\ref{torqueeq}).

\begin{figure}[t!]
\centering
\vspace{-0cm}
\includegraphics[width=1\linewidth]{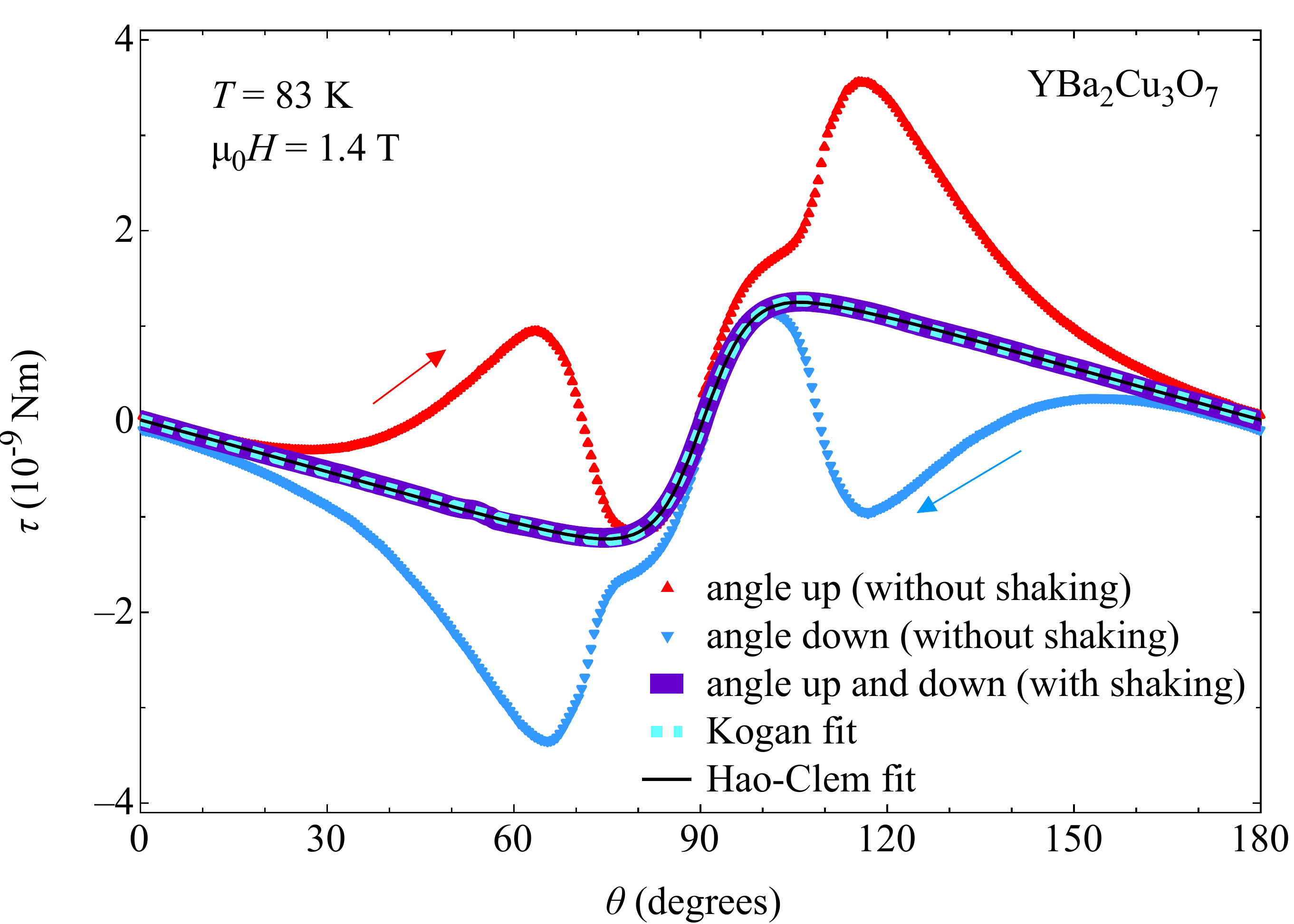}
\caption{(Color online) Magnetic torque $\tau$ as a function of $\theta$ for single crystal YBa$_2$Cu$_3$O$_7$ with and without vortex shaking at 83 K and 1.4 T. The shaking removes the irreversibility between the increasing angle (up) and decreasing angle (down) measurements. The Kogan and the HC models both describe the data equally well. }
\label{shake}
\end{figure}

In general, the torque signal is distorted by pinning effects: the vortex cores are pinned by defects in the sample, in which superconductivity is more easily suppressed. Consequently, the sample is not at thermodynamic equilibrium during the time span of one measurement. As a result, the torque signals are different for angular fields measurements in opposite directions. In order to get reversible angular dependent torque data, a "vortex shaking" technique\cite{Willemin1999} was used. In this technique a small AC field ( $\simeq$ 200 Hz, 1 mT) is applied perpendicular to the main external field in order to shake the vortices out of their pinning sites (Fig.~\ref{shake}).

The temperature range of the angular measurements was $\rm{77~K~to~86~K}$. The lower temperature bound was chosen such as to avoid the lock-in effect (also known as intrinsic pinning),\cite{Feinberg1993} which influences the torque in a way not accounted for in Eq.~(\ref{torqueHC}). When the external field direction becomes close to the $ab$-plane ($\theta= 90$\textdegree), the magnetization abruptly aligns with the $ab$-planes in order to minimize the magnetic energy in the superconducting state. However, it "jumps" back outside the planes when the external field direction is sufficiently away from the $ab$-plane. The upper temperature bound was chosen such as to avoid fluctuation \changeStart effects\cite{Hofer2000} \changeStop close to $T_{\rm c}$. Fluctuation effects are not taken into account in the mean-field approximation of the models considered here.

\begin{figure}[t!]
\centering
\vspace{-0cm}
\includegraphics[width=1\linewidth]{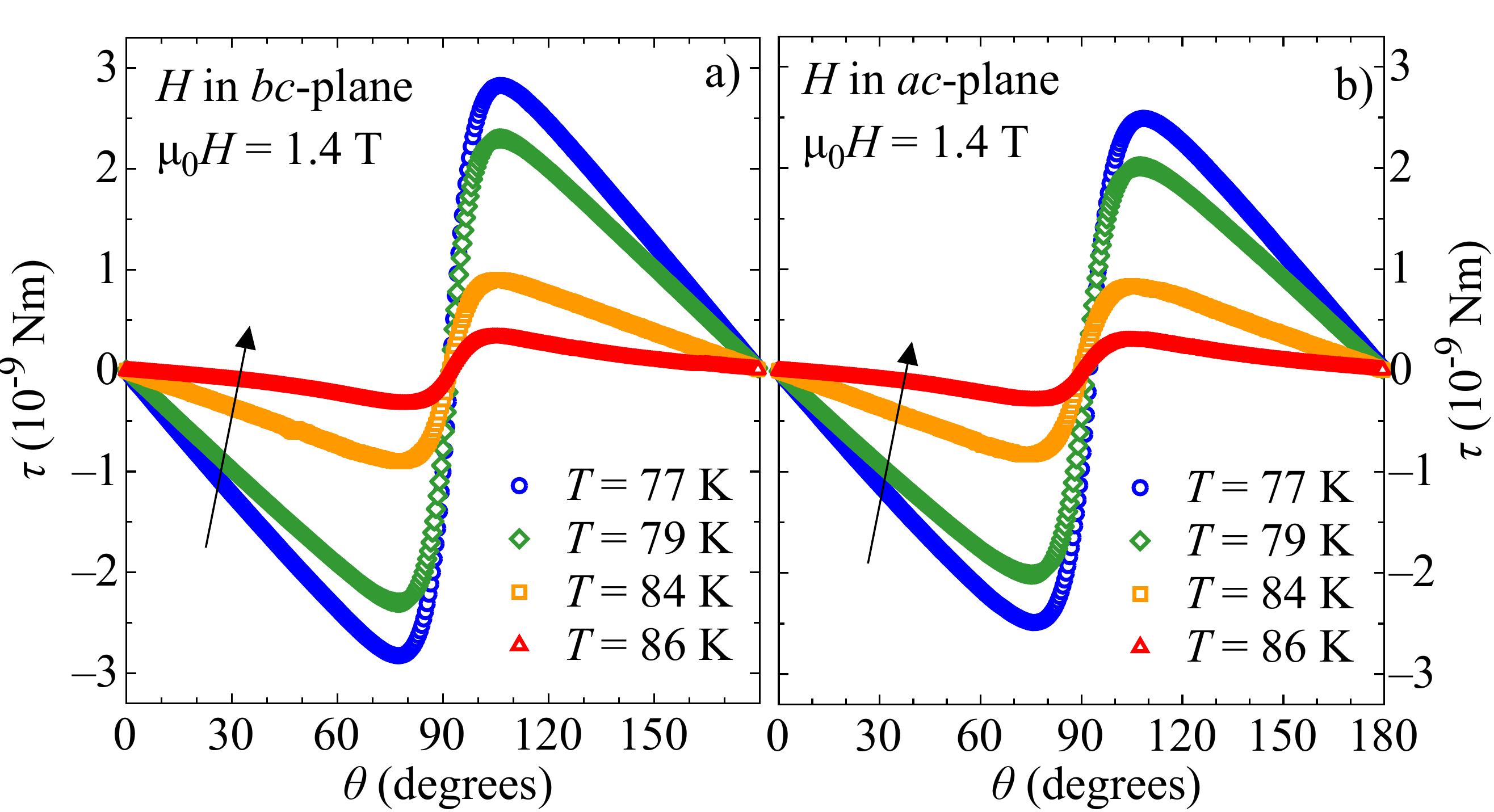}
\caption{(Color online) Angular dependent torque measurements of a single crystal YBa$_2$Cu$_3$O$_7$ taken in the temperature range between 77 K and $T_{\rm c}$ at 1.4 T. (For clarity not all temperatures are shown; the arrows indicate increasing temperature). These raw data include a sinusoidal background. a) Measurements with $H$ in the $bc$-plane. b) Measurements with $H$ in the $ac$-plane.}
\label{torqueACBC}
\end{figure}

YBa$_2$Cu$_3$O$_7$ has an orthorhombic structure. Taking into account that in this case $\lambda_a \neq \lambda_b$, one has to replace $\gamma$ by $\gamma_{ca}$ or $\gamma_{cb}$ in Eqs.~(\ref{ML}), (\ref{torqueL}), (\ref{MHC}), and (\ref{torqueHC}), with the magnetic field direction in the $ac$- or $bc$-plane,\cite{Ager2000,Willemin1999} and $\lambda_{ab} = \sqrt{\lambda_{a}\lambda_{b}}$ is not equal to $\lambda_a$ and $\lambda_b$ as in the tetragonal case. In order to check the validity of this tetragonal approximation, measurements as a function of angle were performed in both the $bc$-plane and $ac$-plane (see Fig.~\ref{torqueACBC}). As expected, the data are similar for both orientations, thus allowing the analysis within a tetragonal model. 
   
\begin{figure}[t!]
\centering
\vspace{-0cm}
\includegraphics[width=1\linewidth]{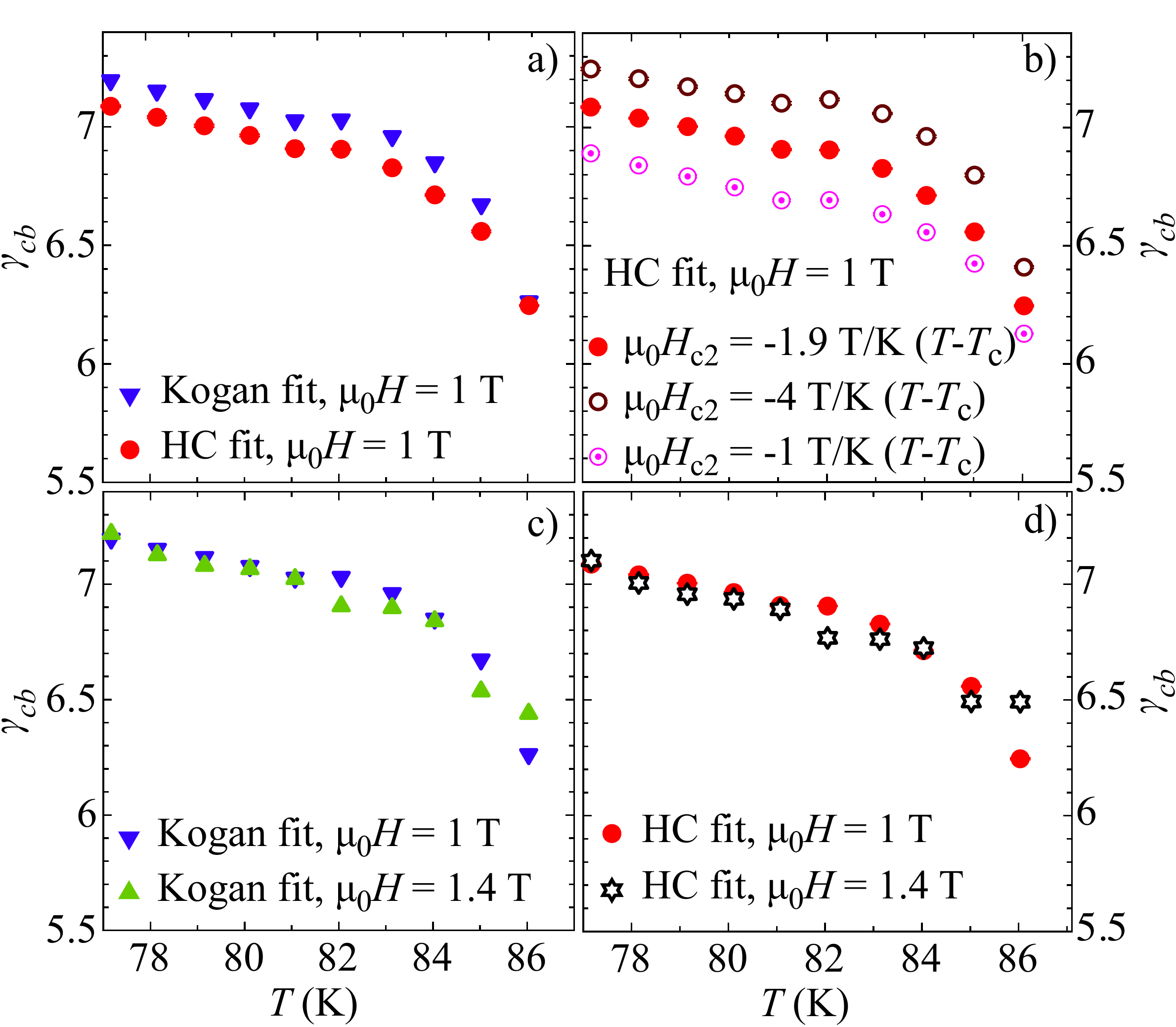}
\caption{(Color online) Overview of the results of the angular dependent magnetic torque measurements of single crystal YBa$_2$Cu$_3$O$_7$ for various temperatures and fields, using the models described in the text. a) Comparison of $\gamma_{cb}$ for the HC and Kogan model at fixed $\mu_0H = 1~\rm{T}$ and fixed upper critical field ($\mu_0 dH^{||c}_{\rm c2}$/$dT = - 1.9~\rm{T/K}$). The Kogan model yields a slightly larger anisotropy parameter than the HC model. b) $\gamma_{cb}$ for the HC model at $\mu_0H = 1~\rm{T}$ and various upper critical fields. The parameter $\mu_0 dH^{||c}_{\rm c2}$/$dT$ does not change the shape of $\gamma(T)$. c) $\gamma_{cb}$ for the Kogan model at $\mu_0H = 1~\rm{T}$ and $\mu_0H = 1.4~\rm{T}$ and fixed upper critical field ($\mu_0 dH^{||c}_{\rm c2}$/$dT = - 1.9~\rm{T/K}$). d) $\gamma_{cb}$ for the HC model, same conditions as in panel c). Panels c) and d) show that the field dependence of $\gamma$ is only marginal.}
\label{gammaComp}  
\end{figure}

The torque data were analyzed with the HC and with the Kogan model. In order to reduce the number of free fit parameters, the upper critical field was fixed in the fitting procedure according to a Werthamer-Helfand-Hohenberg (WHH) temperature dependence\cite{Werthamer1966} suitable for YBa$_2$Cu$_3$O$_7$:\cite{Welp1989} $\mu_0H^{||c}_{\rm c2} \simeq - 1.9~ {\rm T/K} \cdot (T-T_{\rm c})$. The paramagnetic background signal $\chi(VH^2/2)\rm{sin}(2\theta)$ present in the torque data was subtracted using the method described in Refs.~\onlinecite{Balicas2008, Weyeneth2009b}. Figure~\ref{gammaComp} shows the temperature dependence of the anisotropy parameter $\gamma_{cb}$ as determined from the torque data using the two models. As evident in Fig.~\ref{gammaComp}a, both models yield very similar values for $\gamma_{cb}$ (within 2\% accuracy). Moreover, the results depend only weakly on the value taken for $H^{||c}_{\rm c2}$ (Fig.~\ref{gammaComp}b) and on the external field (Figs.~\ref{gammaComp}c and \ref{gammaComp}d). The errors of the fit parameters $\gamma_{ij}$ and $\lambda_{ab}$ were estimated with a Monte-Carlo method: different fits were performed for randomly sampled points within the experimental error of the measured data points. The final values of the parameters $\gamma$ and $\lambda$ were taken as the average values obtained by this procedure, and their errors were defined as twice the standard deviation of these results. The estimated error bars are smaller than the size of the data points.
 
\begin{figure}[t!]
\centering
\vspace{-0cm}
\includegraphics[width=0.8\linewidth]{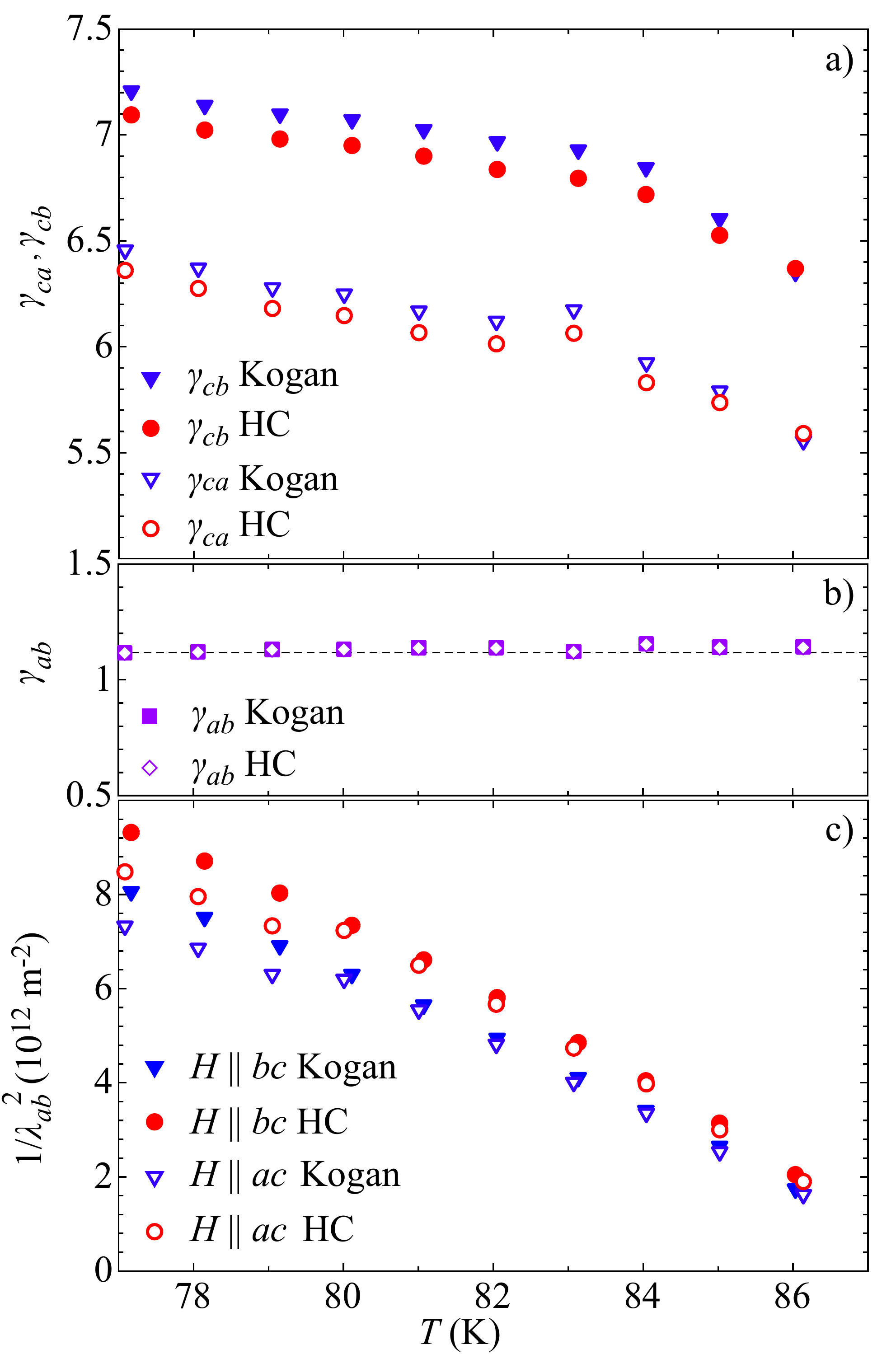}
\caption{(Color online) a) Temperature dependence of the anisotropy parameters $\gamma_{ca}$ and $\gamma_{cb}$ for single crystal YBa$_2$Cu$_3$O$_7$, obtained by averaging $\gamma$ for both measured fields (1 T and 1.4 T). b) Temperature dependence of the in-plane anisotropy parameter $\gamma_{ab}$. The dotted line is the average $\gamma_{ab} = 1.12(5)$. c) Temperature dependence of $1/\lambda_{ab}^{2}$~for measurements with $H$ parallel to the $ac$- and $bc$- planes.} 
\label{gammaT} 
\end{figure}

\changeStart Since the anisotropy parameter is only weakly field dependent (see Figs.~\ref{gammaComp}c and \ref{gammaComp}d), we take as the final $\gamma$ value the average for 1.4 T and 1 T.\changeStop The corresponding temperature dependences of $\gamma_{ca}$ and $\gamma_{cb}$ are shown in Fig.~\ref{gammaT}a. The free fit parameters in Eqs.~(\ref{torqueL}) and (\ref{torqueHC}) are the anisotropy parameter $\gamma_{ij}$ and the in-plane magnetic penetration depth $\lambda_{ab}$. Since the volume $V$ of the sample is not known precisely, the extracted value for $\lambda_{ab}$ may deviate from the intrinsic value. However, the shape of $\lambda_{ab}(T)$ reflects the true temperature dependence, because the volume $V$ is only slightly temperature dependent. Figure~\ref{gammaT}c shows the temperature dependence of  $1/\lambda_{ab}^{2}$ as estimated from the torque data using the Kogan and HC model. Over the temperature range studied, $\gamma_{ca}$ as well as $\gamma_{cb}$ slightly increase with decreasing temperature, whereas the in-plane anisotropy parameter $\gamma_{ab}$ is temperature independent, in fair agreement with previous $\mu$SR measurements of the magnetic penetration depth obtained for a similar sample.\cite{Khasanov2007} However, since $\gamma_{ab} \simeq 1$ one should note that it is difficult to draw definite conclusions about its temperature dependence. The temperature dependence of the magnetic penetration depth from the previous $\mu$SR study\cite{Khasanov2007} was measured along the three principal crystallographic axes and was interpreted in terms of a mixed order parameter of {\it s}+{\it d} wave symmetry. The values of $\gamma_{ca}$, $\gamma_{cb}$ and $\gamma_{ab}$ determined at $T~\simeq~80$~K, together with the values obtained by various experimental techniques at different temperatures are summarized in Table~\ref{gamValues}. The small differences in the values may be due to the different temperature ranges, the experimental techniques used, or slight differences in the doping of the samples. \changeStart A determination of the anisotropy from the ratio of the penetration depths requires a careful evaluation of $\lambda_c$ and $\lambda_{ab}$. Any misalignement of the sample with the applied magnetic field will result in an underestimation of $\lambda_c$ and the deduced $\gamma$. In torque measurements, however, the anisotropy is extracted from a fit to the data, without orientation issues since the model describes the variation of torque in the full angular range. Therefore, the obtained values are much more reliable. \changeStop The vortex shaking technique allows us to avoid an overestimation of the anisotropy \changeStart due \changeStop to pinning.\cite{Willemin1998a} The parameters $\gamma_{ca}$ and $\gamma_{cb}$ are slightly different, because of the orthorhombicity of the crystal structure. The torque data were analyzed here under the assumption that the field and penetration depth anisotropy parameters are equal. It is possible to generalize this analysis to the multi-gap case, where these parameters are not equal.\cite{Kogan2002} However, such an analysis of the present torque data would not provide reliable results due to the too large number of fit parameters.

\begin{table}[ht] \caption{Comparison of anisotropy parameters of YBa$_2$Cu$_3$O$_{7-\delta}$ obtained by various experimental techniques at different temperatures.} 
\renewcommand{\thefootnote}{\thempfootnote}
\centering 
\begin{tabular}{lccccc} 
\hline\hline 
 technique & $T$ (K) & $\gamma_{ab}$ & $\gamma_{cb}$ &  $\gamma_{ca}$ & Ref. \\ 
\hline 
 low-energy $\mu$SR & 0  &  1.19(1) & . &  . & \onlinecite{Kiefl2010} \\
 SANS\footnote{small angle neutron scattering} & 1.5  & 1.18(2) & . & . & \onlinecite{Johnson1999} \\
 $\mu$SR & 10  & 1.15(2) & 4.2(5) & 3.6(4) & \onlinecite{Ager2000} \\
 $\mu$SR & 80  & 1.1(1)\footnote{estimated from Ref.~\onlinecite{Khasanov2007}} & 4.5(1)\footnotemark[\value{mpfootnote}] & 3.5(1)\footnotemark[\value{mpfootnote}] & \onlinecite{Khasanov2007} \\
 specific heat & 70-90\footnote{temperature not specified; out-of-plane anisotropy parameter determined from $H_{c2}$}  & . & 5.3(5) &  5.3(5) & \onlinecite{Roulin1998} \\
 torque \changeStart (shaken)\changeStop & 80  &  1.12(5) & 7.00(5) & 6.18(5) & this work \\
 torque & 90  &  1.18(14) & 8.95(76) & 7.55(63) & \onlinecite{Ishida1996} \\ 
 torque & 93  &  1.1(2) & 7.3(5) &  6.6(5) & \onlinecite{Ager2000} \\
\hline\hline
\end{tabular} 
\label{gamValues} 
\end{table}

Although no temperature dependence of the out-of-plane anisotropy parameter for Pr-doped YBa$_2$Cu$_3$O$_{7-\delta}$ was found,\cite{Kortyka2010} it was noted that such a dependence cannot be ruled out due to the narrow temperature range studied (82 - 88 K). A temperature independent out-of-plane anisotropy parameter was also observed for HgBa$_2$Ca$_3$Cu$_4$O$_{10}$,\cite{Zech1996} where the studied temperature range was very narrow as well. In contrast, a pronounced temperature dependence of the out-of-plane anisotropy parameter was seen in MgB$_2$,\cite{Angst2002} which was consistently described in the framework of two-gap superconductivity. The temperature dependence of the out-of-plane anisotropy parameter found later in iron-based superconductors was also attributed to multi-gap superconductivity.\cite{Weyeneth2009a, Weyeneth2009b} This may suggest that the present results are a signature of two-gap superconductivity in YBa$_2$Cu$_3$O$_7$, as previously proposed in Ref.~\onlinecite{Khasanov2007}. However, we note that the temperature dependence of the out-of-plane anisotropy parameter observed for cuprates is extremely sensitive to the oxygen content. A well pronounced temperature dependence of the anisotropy for strongly underdoped samples\cite{Kortyka2010a} becomes very weak for overdoped YBa$_2$Cu$_3$O$_7$. This may be related to the evolution of the pseudogap with doping in YBa$_2$Cu$_3$O$_{7-\delta}$. Moreover, it suggests that an additional energy scale to the superconducting energy gap in the system is necessary to get a temperature dependent out-of-plane anisotropy parameter in layered superconductors. Such an energy scale may originate from the multi-gap nature of superconductivity in MgB$_2$ and in pnictides and from the appearance of the pseudogap in cuprates.

\section{Conclusions} 

The magnetic torque of an overdoped YBa$_2$Cu$_3$O$_7$ single crystal was investigated at temperatures close to $T_{\rm c}$ in magnetic fields of 1 T and 1.4 T. In the temperature range $0.87~T_{\rm c} < T < T_{\rm c}$, the anisotropy parameters  $\gamma_{ca}$ and $\gamma_{cb}$ were found to increase by more than 10\% with decreasing temperature, but no field dependence was observed. In contrast, the in-plane anisotropy parameter $\gamma_{ab}$ exhibits no temperature nor field dependence. The values of $\gamma_{ca}$, $\gamma_{cb}$ and  $\gamma_{ab}$ are in good agreement with those reported previously (see Table~\ref{gamValues}). The analysis of the torque data with the Hao-Clem model yields within 2\% the same results as the simpler Kogan model. The Hao-Clem model does not provide new information on the vortex state of YBa$_2$Cu$_3$O$_7$ in the present study.

The weak temperature dependence of the out-of-plane anisotropy parameter may indicate the presence of two energy scales in the superconducting behavior, related to multi-gap superconductivity or one-gap superconductivity with a pseudogap. To clarify this hypothesis more experimental work is required.
 
\section*{Acknowledgements}
  
This work was partly supported by the Swiss National Science Foundation.
\\


%

\end{document}